\documentclass[12pt]{elsartapp}

\usepackage{epsfig,graphicx}
\begin{document}
\hyphenation{a-man-da sen-na phys-ics al-iz-ar-in phys-ics}
\sloppy
\begin{frontmatter}

\title{Unbiased cut selection for optimal upper limits in neutrino detectors: the model 
rejection potential technique}

\author{{Gary~C.~Hill}\corauthref{cor}} and 
\corauth[cor]{Corresponding author. Tel.: +1 608 263 1937 fax: +1 608 263 0800}
\ead{ghill@sen\-na.phys\-ics.wisc.edu}
\author{Katherine Rawlins\thanksref{kathsp}}
\thanks[kathsp]{Present address: Amundsen-Scott South Pole Station, Antarctica}
\ead{kath@a\-man\-da.spole.gov}

\address{Department of Physics}
\address{University of Wisconsin, Madison}
\address{1150 University Avenue}
\address{Madison WI 53706 USA}

\begin{abstract}
We present a method for optimising experimental cuts in order to place the
strongest constraints (upper limits) on theoretical signal models.
  The method relies only on signal 
and background expectations derived from Monte-Carlo simulations, so no bias is
introduced by looking at actual data, for instance by setting a limit based on
expected signal above the ``last remaining data event.''
After discussing the concept of the ``average upper limit,'' 
based on the expectation from an ensemble of repeated experiments 
with no true signal,
we show how the best model rejection potential
is achieved by optimising the cuts to 
minimise the ratio of
this ``average upper limit'' to the expected signal from the model.  
As an example, we use this technique to determine the limit sensitivity of kilometre
scale neutrino detectors to extra-terrestrial neutrino fluxes from a variety of
models, e.g. active galaxies and gamma-ray bursts. We suggest that these 
model rejection potential optimised
limits be used as a standard method of comparing the sensitivity of proposed
neutrino detectors.
\end{abstract}
\end{frontmatter}

\clearpage

\section{Introduction}
In this paper, we examine the problem of choosing experimental cuts
in order to 
place the most restrictive limits on
theoretical signal models. 
How should one choose cuts with this in mind?
If a signal is assumed, the goal is to maximise the significance of the
observation, by for instance optimising signal to noise or signal
 to square root noise. 
 If, however, one assumes that no signal will be observed, a 
different technique is required to optimise the ``model rejection potential,''
 i.e. the 
limit setting potential of an experiment. This technique must not depend
 on the experimental data, since choosing cuts based
on the data (e.g. cutting after the last remaining event) leads to confidence 
intervals that do  not have frequentist coverage.  
 
 In this paper, we describe and assess an unbiased
   method~\cite{brussels,hartill}, based only on Monte Carlo signal and background
expectations.

 Firstly we discuss how an experimental observation is
used to set a limit 
on a theoretical flux (section \ref{howtolimits}).
 The desire to minimise the upper limit (and thus place
the strongest constraint on the theoretical model) leads to the concepts
of ``average upper limits'' and ``model rejection factor,'' which are discussed
 in section 
\ref{aulamrp}. We show how choosing cuts based on optimising the 
``model rejection factor'' leads to, on average, 
 the best possible limit from the
experiment.
In section \ref{km3sens}, we illustrate this technique  
by calculating the model rejection potential of a
kilometre  scale neutrino detector with respect to predicted 
extra-terrestial neutrino fluxes such as active galaxies and gamma-ray bursts.

\section{Calculating flux limits}
\label{howtolimits}

When an experiment fails to detect an expected flux, an upper limit on that flux
can be derived from the experimental observation. As an example consider a detector
that should observe 100 signal events above an expected background of a few events. If
only the few events consistent with background 
are observed, then obviously the model is strongly constrained as
the 100 expected events were not seen. 
 The exact calculation of an upper limit on the
source flux would proceed as follows.
The theoretical source flux, $\Phi(E,\theta)$, is
convolved with the detector response and after some cuts yields an event expectation,
$n_s$. Let the expected background for the same analysis and cuts be $n_{b}$. The
experiment is performed and $n_{\mathit{obs}}$ events are seen. 

In this experiment the 90\% confidence\footnote{For
simplicity we will
only use 90\% confidence intervals in this discussion; the results of course are
applicable to any level of confidence or any formulation of confidence intervals -- whether 
frequentist or Bayesian.}
 interval $\mu_{90}=(\mu_1,\mu_2)$
 is a function of the number of
observed events, $n_{\mathit{obs}}$, and of the expected background $n_{b}$
\begin{equation}
       \mu_{90}(n_{\mathit{obs}},n_{b})
\end{equation}
For example, an experiment that observes 3 events on an expected background of
1.5 would report a $90\%$ confidence interval in the Feldman-Cousins~\cite{fc98}
 ``unified'' approach
of $\mu_{90}(3,1.5) = (0.0,5.92)$. Note that this confidence interval 
includes 0.0, and  
therefore $\mu_2 = 5.92$ is an upper limit. In what follows
 we shall take $\mu_{90}$ to mean this upper limit. 

The corresponding upper limit 
on the source spectrum $\Phi(E,\theta)$ is found by scaling
the source flux by the ratio of the upper limit to the signal expectation~\cite{frejus_diffuse} 
\begin{equation}
\label{scaleflux}
   \Phi(E,\theta)_{90\%} = \Phi(E,\theta) \frac{\mu_{90}(n_{\mathit{obs}},n_{b})}{n_s}
\end{equation}
If the expected signal contribution to the observation was $n_s = 100$, then the $90\%$
confidence level upper limit on the source flux is 
\begin{equation} 
      \Phi(E,\theta)_{90\%} = \Phi(E,\theta) \times \frac{5.92}{100}
\end{equation}
In this case the theoretical model predicts
 a large number of events with very
few 
 seen and is therefore severely constrained by the experiment. A low value 
of the ratio $\mu_{90}(n_{\mathit{obs}},n_{b})$/$n_s$ has led to a strong constaint on the
model, therefore  
one should optimise cuts to make this ratio
small.  However, the observed upper limit depends on $n_{\mathit{obs}}$ which is not known
until the cuts are defined and the experiment performed! Fortunately, the classical 
concept of an ensemble of experiments allows us to calculate an ``average upper limit'' 
(equivalent to the Feldman-Cousins ``sensitivity''~\cite{fc98})
which can take the place of $\mu_{90}(n_{\mathit{obs}},n_{b})$ in the signal to upper limit
calculation.

\section{Average upper limits and ``model rejection factor''}
\label{aulamrp}

Although we cannot know the actual upper limit that will result from an experiment
until looking at the data, we can use the Monte Carlo predictions to 
 calculate the average upper limit (Feldman-Cousins ``sensitivity''~\cite{fc98})
 that would be observed after 
hypothetical repetition of the experiment with 
expected background $n_{b}$ and no true signal ($n_s=0$).
In the example of the previous section,
 the expected background was $n_{b}=1.5$. Over an ensemble of
experiments with no true signal,
  this background will fluctuate
 to $n_{\mathit{obs}}=0$ with Poisson probability 22.3\% and an upper limit of 
$\mu_{90}(0,1.5)=1.33$ reported. 
The background will fluctuate to $n_{\mathit{obs}}=1$ in 33.5\% of experiments, where
 an upper limit
of 2.91 would be reported, and so on for each value of $n_{\mathit{obs}}$. 
The ``average upper limit''  is the
sum of these expected upper limits, weighted by their Poisson 
probability of occurrence
\begin{equation}
\label{aul}
        {\bar{{\mu}}_{90}(n_{b})} = \sum_{n_{\mathit{obs}}=0}^{\infty} 
     \mu_{90}(n_{\mathit{obs}},n_{b}) \;\; \frac{(n_{b})^{n_{\mathit{obs}}}}{(n_{\mathit{obs}})!} \exp(-n_{b})
\end{equation}

\begin{figure}[htp]
\centering
 \mbox{\epsfig{file=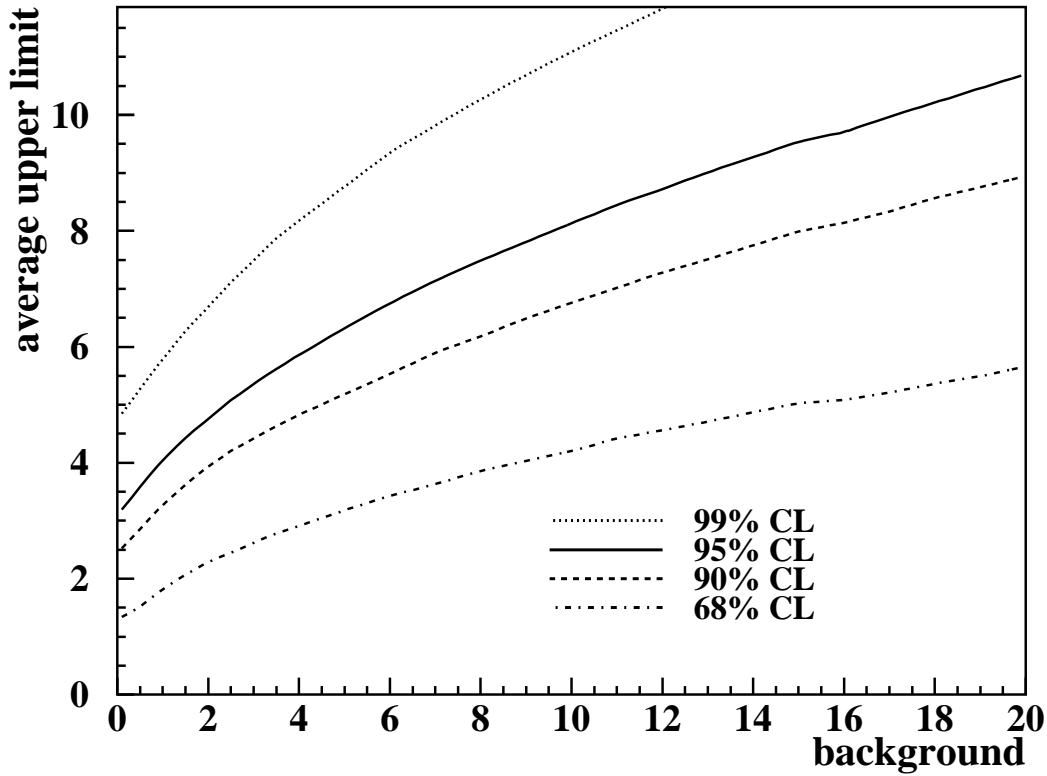,width=15cm}}
\caption[1]{
\label{cflimit}  Average 90\% Feldman-Cousins upper limit that would be
obtained by an ensemble of
experiments with no true signal in the presence of expected
background. 
 }
\end{figure}

This average upper limit is shown in figure~\ref{cflimit} as a function of
the expected background for four different Poisson confidence levels under 
  the Feldman-Cousins ``unified'' ordering principle.
 Before performing the experiment, one can
  see what background
 would be expected to remain as a function of various cuts,
then
consult this plot to find out what average upper limit 
result the experiment would
be expected to give.

Over an ensemble of identical experiments, the 
strongest constraint on the expected signal flux $\Phi(E,\theta)$
corresponds to the set of cuts that  minimises the 
``model rejection factor''
\begin{equation} 
              \frac{\bar{\mu}_{90}}{n_s}
\end{equation}
and hence minimises the average flux upper limit that would be obtained 
over the hypothetical 
experimental ensemble
\begin{equation}
           \bar{\Phi}(E,\theta)_{90} =  \Phi(E,\theta) 
             \frac{\bar{\mu}_{90}}{n_s}
\end{equation}

In the actual experiment we will not obtain $\bar{\Phi}(E,\theta)_{90}$, but obtain one
limit based on the  observed number of background counts, which depends on how
the background happened to fluctuate in the single, real experiment. 
However, prior to performing the experiment, the average flux upper limit  (flux times 
model rejection factor) tells us
what we would expect over repeated runs of the real experiment, and on average the
best limit will come from choosing cuts that minimise this average limit. 
Also, the model rejection factor curve tells us how sensitive the expected
limit is
to the choice of cuts.

A somewhat common practice is to choose the final cut based on the last observed
data event. The event limit $\mu_{90}$ then comes from the observation of zero events above that
cut. This event limit is then 
 compared to the predicted signal from the source flux above the cut, and the flux limit obtained
using equation \ref{scaleflux}. Calculating the event limit in this way
results in confidence intervals which do not have frequentist coverage. The
classical concept of frequentist coverage~\cite{Neyman,fc98,Cousins94} involves the
construction of $(100\times\alpha)\%$  confidence intervals, such that
 in $(100\times\alpha)\%$ of repeated
experiments, the reported confidence interval will include (``cover'') the fixed
but unknown true value of the estimated parameter $\mu_{t}$. This coverage 
applies to any value that $\mu_{t}$ may take. Choosing the cut based on the
last observed data event will lead to intervals that do not cover all possible
values of  $\mu_{t}$ with the correct frequency, as
 can be seen by a simple example. Assume that we are 
using a standard Neyman construction for a Poissonian upper limit. As a example, 
the 90\% 
confidence level upper
limit for zero observed events is $\mu = 2.3$. If we always choose the cut such that 
zero events are observed, then we will always report an upper limit $\mu = 2.3$.
 Thus, in any experiments
where the true value of the parameter $\mu$  is greater than 2.3, the reported
 confidence interval, (always $\mu < 2.3$), never covers the true value! At the
other extreme, for
true values $\mu_t < 2.3$, the reported confidence interval will
 cover the true value in
exactly $100\%$ of cases. In neither case is the required 90\% coverage achieved, 
showing that choosing the final cut based on the observed data is not a 
 valid method in the frequentist framework.

 These considerations show that the final experimental
 cuts must be chosen prior to looking at the 
data, and that the model rejection potential technique provides a method of optimising
 these cuts to give the best average limit.
 In the next section we apply the model rejection potential
 technique to the 
determination of the sensitivity of proposed kilometre scale neutrino
detectors to astrophysical sources of neutrinos, using searches for both
diffuse and point like emissions as examples.

\section{Model rejection potential of kilometre scale neutrino detectors}
\label{km3sens}
Various proposals exist for the construction of kilometre scale neutrino 
detectors~\cite{icrc2001} in the the deep ocean
 (the ANTARES, NESTOR and NEMO experiments) or in Antarctic 
ice (the Icecube experiment). These experiments look for upward moving muons which 
result from neutrino
interactions in the surrounding media (water, ice or rock). The upgoing requirement
is necessary to reject the large background flux of downgoing cosmic ray induced
muons. After rejecting all downward events, the flux of atmospheric neutrinos from
cosmic ray interactions in the earth's atmosphere remains as the background to
a search for extra-terrestrial neutrinos. 

We can determine the sensitivity of an experiment to a flux of extra-terrestrial
neutrinos by optimising the
model rejection potential to find the 
average limit that would be placed on that flux if 
no true signal were present, and only the expected
background atmospheric neutrino events observed. In the following sections, we
use the searches for  diffuse fluxes and point sources
 of extra-terrestrial neutrinos to illustrate the model rejection potential 
technique. For the diffuse flux case, the model rejection potential
 optimisation is done in
one cut dimension, whereas in the point source case, we show how a two-dimensional
optimisation is made. The method could in principle be generalised to an
optimisation in $n$ dimensions, where the point in the $n$-dimensional cut parameter
space is found where the model rejection factor is minimised.  

Predicting the rate of upward moving neutrino induced muons in the vicinity
of an underwater or underice detector is 
 straightforward~\cite{ZHV92,GHS95,FMR95,FMRradio,GQRS95,MyPhD,gch97,GQRS98,ALS} and
so we  describe  here only the basic elements
of the calculation. The current calculation follows a previously 
 described method~\cite{MyPhD,gch97}, 
where the neutrino cross sections 
are calculated using the ``G'' parton distribution set of 
Martin, Roberts and Stirling~\cite{MRSG}, the Lipari-Stanev~\cite{SL91}
muon propagation
code is used to track the muons to the detector, and the structure   
of the earth is accounted for using the Preliminary Reference Earth Model~\cite{PREM}.
An important difference between this and
 other work is the exact treatment of neutral current regeneration
 of neutrinos in the earth 
using a recursive importance sampling interaction algorithm~\cite{MyPhD,gch97}.
 Neutrino and anti-neutrino events
were generated using an $E^{-1}$ spectrum, and then re-weighted to various astrophysical
flux predictions.  A constant detector muon 
effective area of $1\rm{km}^{2}$ was assumed for all angles and
energies.\footnote{The results of a complete simulation of a realistic detector will be 
forthcoming from the Icecube collaboration\cite{IC}.}        

\subsection{Model rejection potential optimisation for diffuse flux searches}

We consider three types of diffuse neutrino flux predictions, firstly $E^{-2}$
spectra, an example of which is the Waxman-Bahcall diffuse upper bound~\cite{WBUB}
  on the flux 
of astrophysical neutrinos. They calculate that the diffuse flux of  neutrinos
should not exceed a level 
\begin{equation}
 E^{2} \Phi_{\mathrm{WBUB}} = 3.0 \times 10^{-8}\; \mathrm{GeV}\; \mathrm{cm}^{-2}\;
\mathrm{s}^{-1}\; \mathrm{sr}^{-1}
\end{equation} 
 Mannheim, Protheroe and Rachen~\cite{MPR98}
 question the 
assumptions made in the  Waxman-Bahcall upper bound calculation and have computed
fluxes of diffuse neutrinos that violate this upper bound. 
One of the Mannheim, Protheroe and Rachen models is also tested here.
Finally, we determine the sensitivity of a kilometre scale detector to the original
  Stecker, Done, Salamon and Sommers~\cite{SDSS91} diffuse neutrino flux
prediction. The atmospheric neutrino background to these searches 
 is taken from the calculation of
 Lipari \cite{Lipari93}. In calculating this background, 
we have not considered the uncertain contribution from the prompt charm decay
 channel \cite{costa2001,gaisserhonda2002}. 
Including this neutrino flux in a
background estimate\footnote{For point source searches, the
background is {\em measured} off source, in which case an uncertain theoretical
prediction is not important.} requires incorporating the systematics into the limit
calculation\cite{ch92,c02}). Or, one could consider this flux as a source on which
limits can be placed, though difficult to separate from an 
extra-terrestrial source~\cite{gch97}.

\begin{figure}[htp]
\centering
 \mbox{\epsfig{file=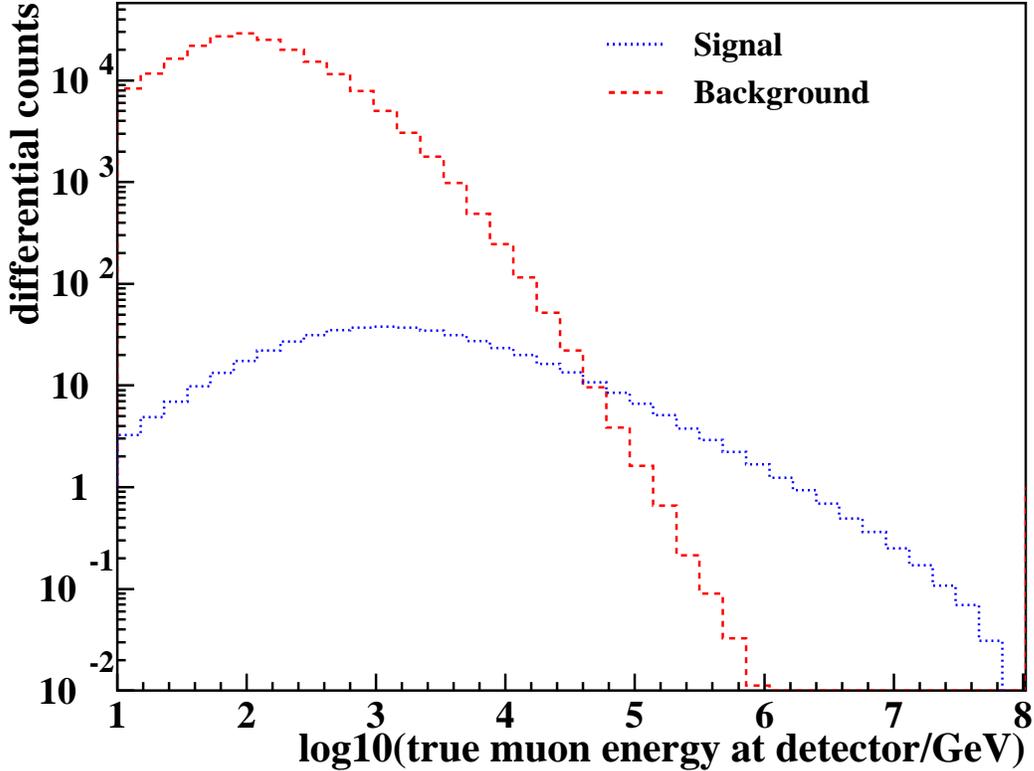,width=15cm}}
\caption[1]{
\label{diffuse_diff}Differential energy distribution of true muon energies
in a neutrino detector for  the atmospheric neutrino flux and an $E^{-2}$
diffuse neutrino flux, of level $E^{2} \Phi_{90} = 3 \times 10^{-8}\; \mathrm{GeV}\; \mathrm{cm}^{-2}\;
\mathrm{s}^{-1}\; \mathrm{sr}^{-1}$, corresponding to the Waxman-Bahcall upper
bound calculation. 
 }
\end{figure}

We consider in detail the optimisation of the model rejection potential for a diffuse
$E^{-2}$ source of astrophysical neutrinos. We will assume that the muon energy
at the detector can be measured, and use this as the cut against the atmospheric
neutrino induced background events. Figure~ \ref{diffuse_diff} shows the differential energy
distributions for the atmospheric neutrino-induced muons, and for the muons
resulting from a  diffuse $E^{-2}$ neutrino source of
strength equal to the Waxman-Bahcall upper bound.

\begin{figure}[htp]
\centering
 \mbox{\epsfig{file=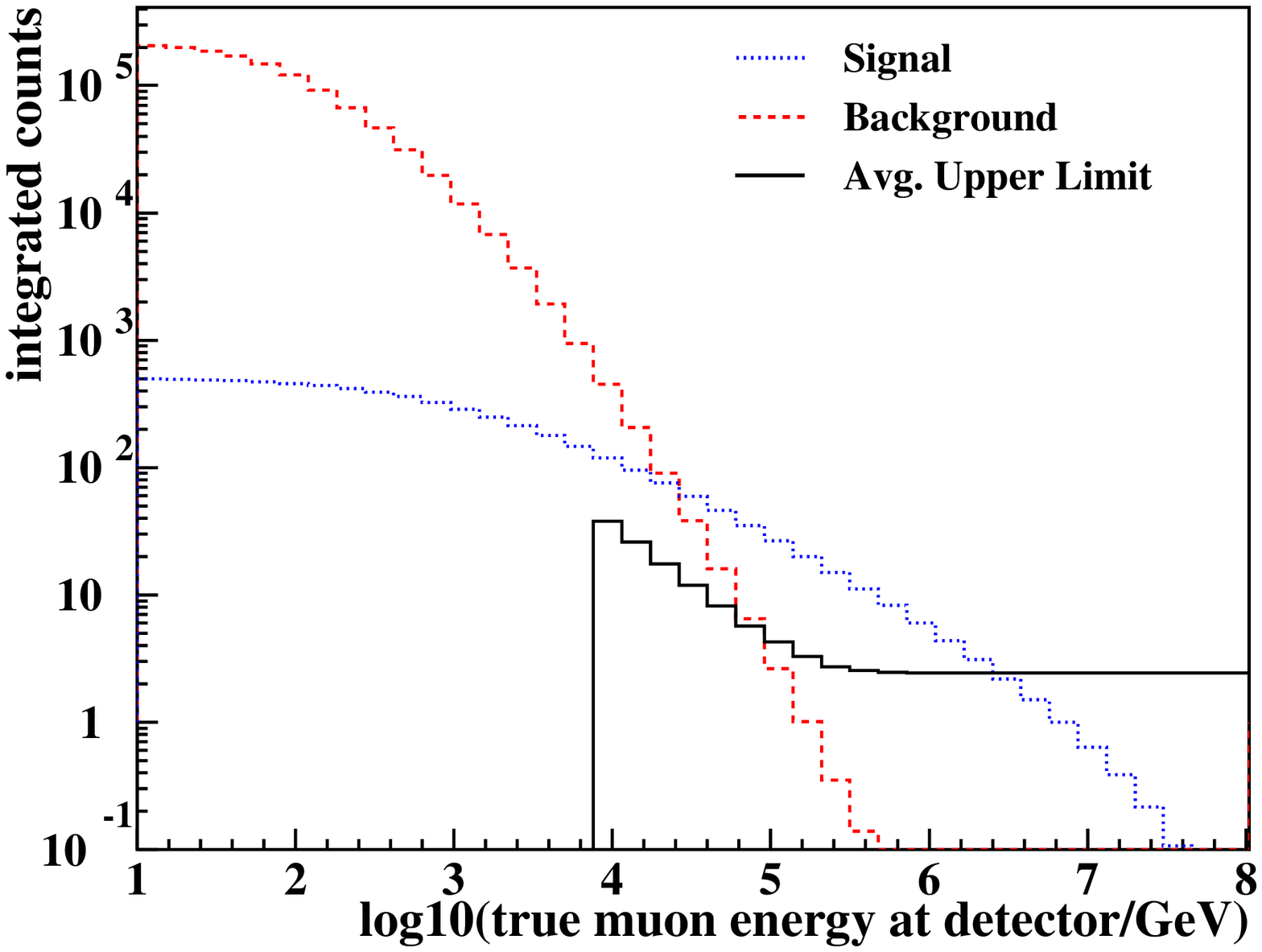,width=15cm}}
\caption[1]{
\label{diffuse_int}Integrated energy distribution of true muon energies
for the fluxes shown in figure~\ref{diffuse_diff}. The average upper  
limit that would be obtained by an ensemble of experiments with no true
signal and only the expected atmospheric neutrino
 background is shown by the solid line. If the background exceeds 700, we do not
calculate an average upper limit, thus the cut off in the curve at $\log_{10}(E_{\mu})=3.87$.
 }
\end{figure}

 Figure~\ref{diffuse_int} shows
the integrated event rates above each muon energy.
 For each value of the expected atmospheric neutrino background, we
can calculate the average upper limit that would be obtained by the 
experimental ensemble with no true signal, and this curve is added to the
plot as a solid line. 

\begin{figure}[htp]
\centering
 \mbox{\epsfig{file=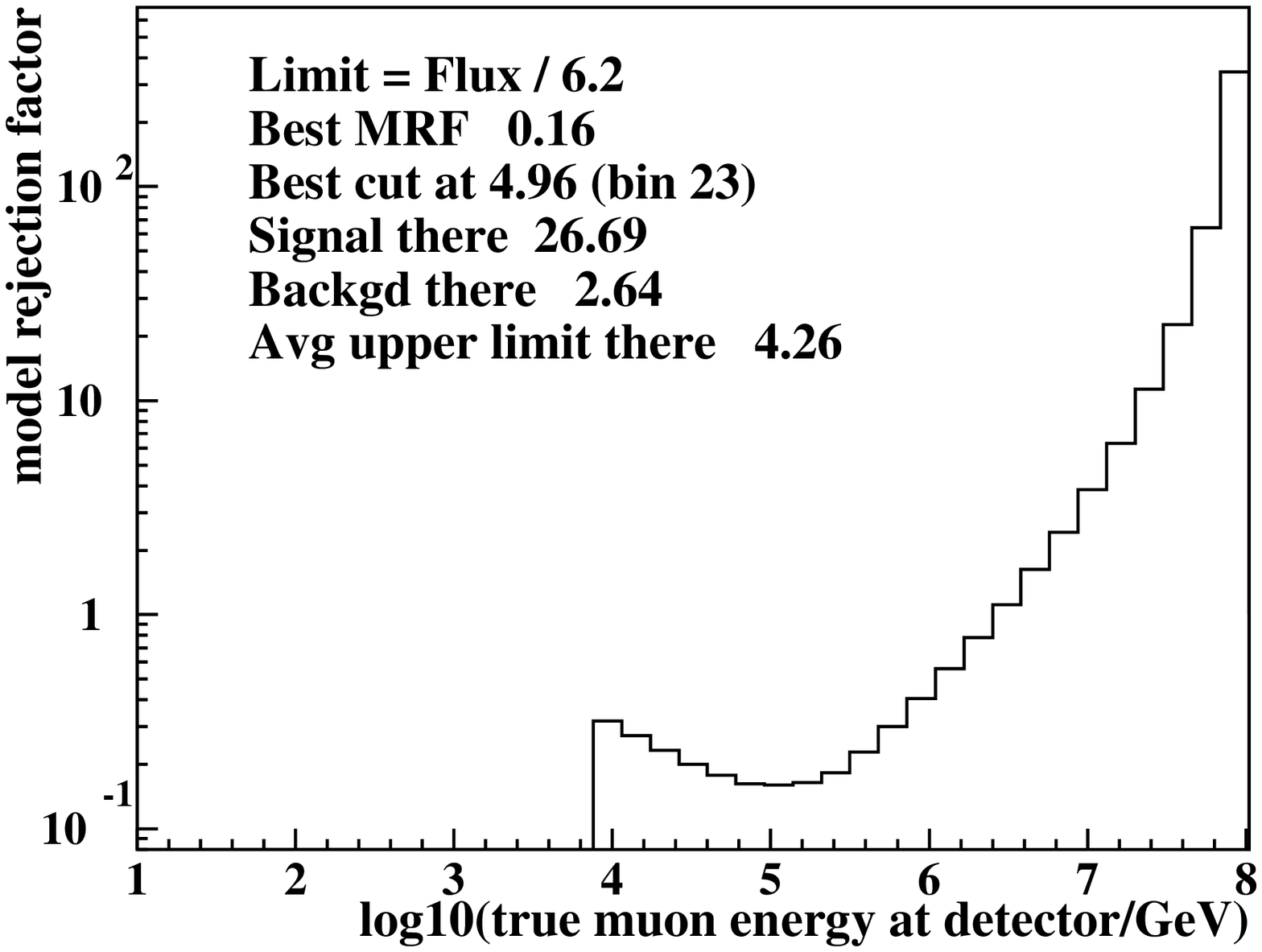,width=15cm}}
\caption[1]{
\label{diffuse_mrf}Model rejection factor, $\bar{\mu}/n_s$, for 
a diffuse flux search in a kilometre effective area neutrino telescope, as a 
function of the muon energy cut. The best average limit on the source flux 
occurs where the MRF is minimised, here at a muon energy just below
 $10^5$ GeV.
 }
\end{figure}

As the background decreases the average upper limit reduces toward 2.44,
the value of $\mu_{90}(n_{\mathit{obs}},n_b)$
 for $n_b=0$ and $n_{\mathit{obs}}=0$. Clearly, if $n_b=0$, the only non-zero
term in the sum in equation \ref{aul} is  
   $\mu_{90}(0,0) = 2.44$ which occurs with probability one (if the background
is zero, then the only possible observation in the absence of signal
 is $n_{\mathit{obs}}=0$).
 The best cut corresponds to  the muon energy
above which the ratio of the average upper limit to the expected
 diffuse signal  is
minimised. This is shown in figure~\ref{diffuse_mrf}.
The best cut lies close to a muon energy of $10^{5}$ GeV, where 2.64 background 
and 26.69 
signal events remain. The
average upper limit for this expected background
 is 4.26, leading to a model rejection factor for the assumed flux strength
of $4.26/26.69 = 0.16$, and thus a limit on such an $E^{-2}$ flux of 
\begin{equation}
 E^{2} \Phi_{90} < 4.8 \times 10^{-9}\; \mathrm{GeV}\; \mathrm{cm}^{-2}\;
\mathrm{s}^{-1}\; \mathrm{sr}^{-1}
\end{equation}
This level of sensitivity is nearly an order of magnitude below that of the Waxman-Bahcall
bound.

Note that this procedure is independent of the assumed signal strength, depending only the
shape of the signal spectrum.

This limit is an ideal one where we have assumed we can measure the muon 
energy exactly.
A realistic detector will have a finite energy resolution,
 and we incorporate this by assuming
a  gaussian resolution in the logarithm of the true muon energy.
 Table \ref{mrpdiffuse1yr} shows a 
summary of the model rejection potential optimisations for measurements of
the true muon energy, for  several assumed detector muon energy resolutions,
 and for measurements
of the true neutrino energy, which would indicate the best possible limit that
 the detector could
achieve. Also, we show what constraints would be placed on other models of neutrino
production, using as examples the calculation of Stecker, Done, Salamon and Sommers~\cite{SDSS91} 
(SDSS), 
and one of the diffuse bounds of Mannheim, Protheroe and Rachen~\cite{MPR98} (MPR). 
A logarithmic muon energy resolution of 30\% produces an expected limit below the 
Waxman-Bahcall upper bound, and is sufficient to place strong constraints on both the SDSS and
MPR models.

\begin{figure}[htp]
\centering
 \mbox{\epsfig{file=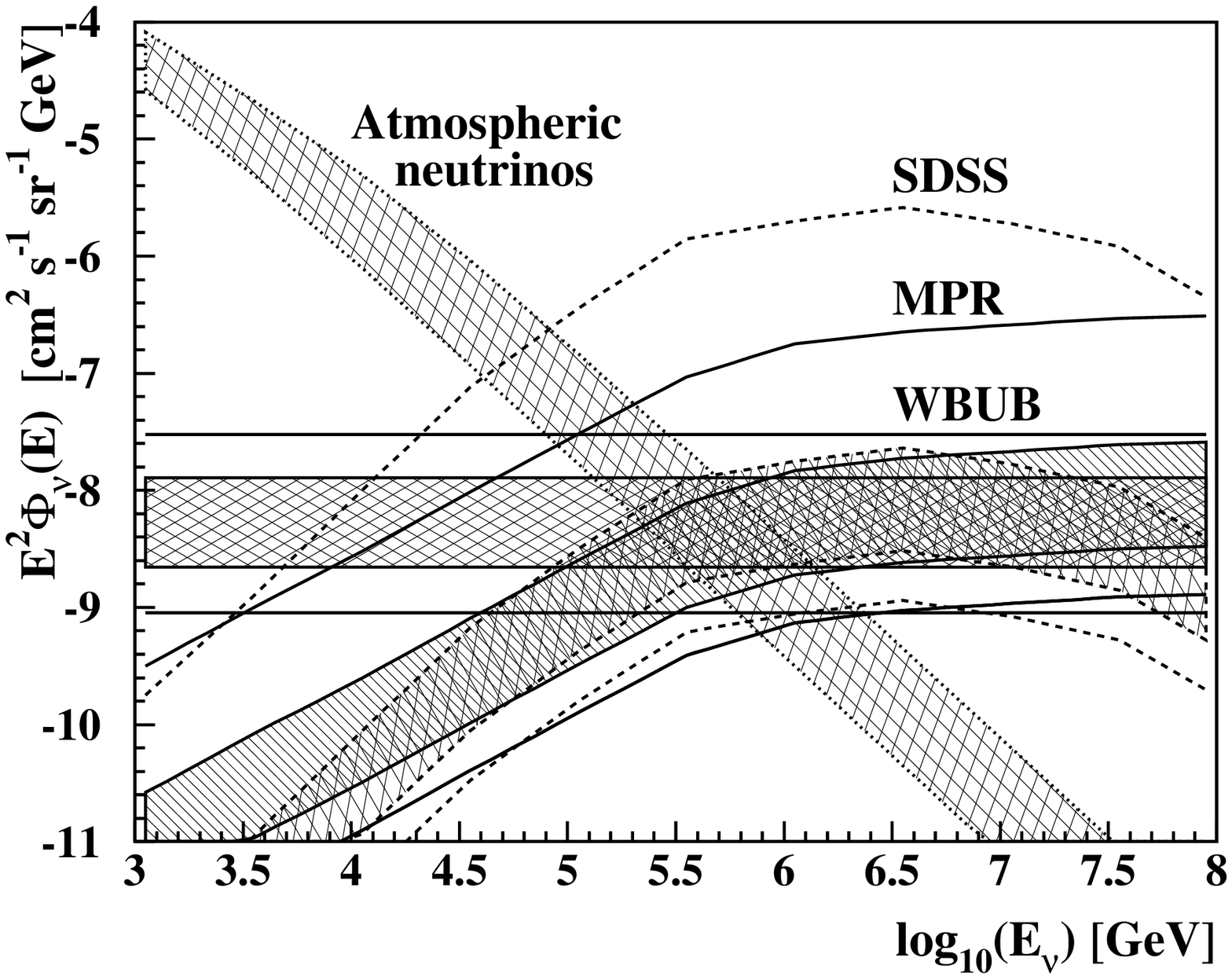,width=15cm}}
\caption[1]{
\label{diffuse_mrf_plot_summary}Summary of the limit setting potential of a kilometre-scale
neutrino detector after three years of operation. For each of the diffuse flux models under
consideration, the highest curves show the predicted neutrino flux. The shaded bands correspond to
the expected range of the limit for different assumptions of the muon energy resolution of the
detector, the upper edge of the band corresponding to 30\% logarithmic muon energy resolution,
 and the lower edge, perfect muon energy resolution. The lowest
curves for each flux show the expected limits if the neutrino energy could be measured directly. The
atmospheric neutrino background is also shown as a band representing the flux variation
 from horizontal (top of band) to  vertical (bottom of band) arrival directions.
 }
\end{figure}

The effects of increasing the detector exposure time are shown in table 
\ref{mrpdiffuse3yr}, where three years of livetime exposure is assumed. In each case
the optimisation moves toward higher energy cut values, and the
improvement in the expected limits goes  approximately as the square root of the 
exposure time. The detector performance for three years exposure is also
summarised
 in figure \ref{diffuse_mrf_plot_summary}.

\begin{table}[htbp]
\caption{\label{mrpdiffuse1yr}Model rejection potential results for various astrophysical
neutrino spectra, for one year live time in a kilometre scale neutrino detector. The Waxman-Bahcall
flux limits are in units $\mathrm{GeV}\; \mathrm{cm}^{-2}\;
\mathrm{s}^{-1}\; \mathrm{sr}^{-1}$.
 }
\vspace{0.5cm}

\begin{tabular}{|c||c|c||c|c|c|c|} \hline

\hline\hline
 &  \multicolumn{2}{|c||}{cut}  & \multicolumn{4}{|c|}{ above cut}\\ 
\hline
Flux & var.($\log$ err.) &  $\log_{10}E$ & $n_s$ & $n_b$
    & $\bar{\mu}(n_b)$ &Flux limit ($=\Phi \times\frac{\bar{\mu}}{n_s}$)\\

    \hline
      \hline
    &  $E_{\nu} (0\%)$ & 5.68 & 51.12 & 1.25 & 3.45 & $2.03\;.\;10^{-9}$\\
    &  $E_{\mu} (0\%)$ & 4.96& 26.69 & 2.64 & 4.26 & $4.79\;.\;10^{-9}$  \\
WBUB    &  $E_{\mu} (10\%) $     & 5.32 & 18.77  & 4.12 & 4.87 & $7.79\;.\;10^{-9}$ \\
    & $E_{\mu} (20\%) $     & 5.68 & 18.47 & 22.19 & 9.37 & $1.52\;.\;10^{-8}$ \\
    &  $E_{\mu} (30\%)$& 6.40 & 14.09 & 32.15 & 10.98  & $2.35\;.\;10^{-8}$\\
    \hline
        \hline
    &  $E_{\nu} (0\%)$ & 5.86 & 281.40 & 0.47 & 2.83 &$\Phi_{\mathrm{MPR}} \times 1.01\;.\;10^{-2}$\\
    &  $E_{\mu} (0\%)$ &5.14 &127.58& 1.01 & 3.29 &$\Phi_{\mathrm{MPR}} \times 2.58\;.\;10^{-2}$  \\
MPR    & $E_{\mu} (10\%)$ & 5.50 & 90.10 & 1.99 & 3.94 &$\Phi_{\mathrm{MPR}} \times 4.37\;.\;10^{-2}$ \\
    & $E_{\mu} (20\%)$     & 6.22 & 54.55 & 4.51 & 5.01 &$\Phi_{\mathrm{MPR}} \times 9.18\;.\;10^{-2}$ \\
    & $E_{\mu} (30\%)$ & 6.94 & 40.53 & 9.09 & 6.51  &$\Phi_{\mathrm{MPR}} \times 0.161$\\
    \hline
       \hline
   &   $E_{\nu}(0\%)$ & 5.68 & 3347.49 & 1.25 & 3.45 &$\Phi_{\mathrm{SDSS}}\times 1.03\;.\;10^{-3}$\\
   &   $E_{\mu} (0\%)$ & 5.14& 1222.01 & 1.01 & 3.29 & $\Phi_{\mathrm{SDSS}}\times 2.69\;.\;10^{-3}$  \\
SDSS   &  $E_{\mu} (10\%)$  & 5.32 & 1039.62  & 4.12 & 4.87 & $\Phi_{\mathrm{SDSS}}\times 4.69\;.\;10^{-3}$ \\
   &  $E_{\mu} (20\%)$ & 6.04 & 606.97 & 7.40 & 6.01 & $\Phi_{\mathrm{SDSS}}\times 9.91\;.\;10^{-3}$ \\
   &  $E_{\mu} (30\%)$ & 6.76  & 452.29& 14.07 &7.77  & $\Phi_{\mathrm{SDSS}}\times 1.72\;.\;10^{-2}$\\
     \hline
\end{tabular}
\end{table}
       
\begin{table}[htbp]
\caption{\label{mrpdiffuse3yr}Model rejection potential results for various astrophysical
neutrino spectra, for three years live time in a kilometre scale neutrino detector. The Waxman-Bahcall
flux limits are in units $\mathrm{GeV}\; \mathrm{cm}^{-2}\;
\mathrm{s}^{-1}\; \mathrm{sr}^{-1}$.
 }
\vspace{0.5cm}

\begin{tabular}{|c||c|c||c|c|c|c|} \hline

\hline\hline
 &  \multicolumn{2}{|c||}{cut}  & \multicolumn{4}{|c|}{ above cut}\\ 
\hline
Flux & var.($\log$ err.) &  $\log_{10}E$ & $n_s$ & $n_b$
    & $\bar{\mu}(n_b)$ &Flux limit ($=\Phi \times\frac{\bar{\mu}}{n_s}$)\\

    \hline

    &  $E_{\nu}$ (0\%)& 5.86 &  118.56 & 1.40 & 3.55 & $8.99\;.\;10^{-10}$\\
    &  $E_{\mu} (0\%)$ & 5.14&  60.21 & 3.04 & 4.43 & $2.21\;.\;10^{-9}$  \\
WBUB    &  $E_{\mu} (10\%)$& 5.50 &  43.12 & 6.01 & 5.53 & $3.85\;.\;10^{-9}$ \\
    &  $E_{\mu} (20\%)$ & 6.04 &  35.48 & 22.06 & 9.34 & $7.90\;.\;10^{-9}$ \\
    & $E_{\mu} (30\%)$& 6.58 &  34.73 & 62.20 & 14.82  & $1.28\;.\;10^{-8}$\\
    \hline
        \hline
    &  $E_{\nu} (0\%)$ & 6.04 & 690.04 & 0.52 & 2.88 &$\Phi_{\mathrm{MPR}} \times 4.17\;.\;10^{-3}$\\
    &  $E_{\mu} (0\%)$ &5.32 &311.16& 1.06 & 3.32 &$\Phi_{\mathrm{MPR}} \times 1.07\;.\;10^{-2}$  \\
MPR    & $E_{\mu} (10\%)$ & 5.68 & 219.77 & 2.84 & 4.35 &$\Phi_{\mathrm{MPR}} \times 1.98\;.\;10^{-2}$ \\
    &  $E_{\mu} (20\%)$   & 6.58 & 112.87 & 4.33 & 4.94 &$\Phi_{\mathrm{MPR}} \times 4.38\;.\;10^{-2}$ \\
    &  $E_{\mu} (30\%)$ & 7.12 & 103.89 & 18.42 & 8.65  &$\Phi_{\mathrm{MPR}} \times 8.32\;.\;10^{-2} $\\
    \hline
       \hline
   &   $E_{\nu} (0\%)$ & 5.86 & 8076.81 & 1.40 & 3.55 &$\Phi_{\mathrm{SDSS}}\times 4.40\;.\;10^{-4}$\\
   &   $E_{\mu} (0\%)$ & 5.32& 2841.39 & 1.06 & 3.32 & $\Phi_{\mathrm{SDSS}}\times 1.17\;.\;10^{-3}$  \\
SDSS   &   $E_{\mu} (10\%)$  & 5.50 & 2481.94  & 5.96 & 5.52 & $\Phi_{\mathrm{SDSS}}\times 2.22\;.\;10^{-3}$ \\
   &  $E_{\mu} (20\%)$  & 6.40 & 1217.47 & 7.67 & 6.09 & $\Phi_{\mathrm{SDSS}}\times 5.00\;.\;10^{-3}$ \\
   &  $E_{\mu} (30\%)$ &7.12  & 967.18& 17.89 &8.54  & $\Phi_{\mathrm{SDSS}}\times 8.83\;.\;10^{-3}$\\
\hline
\end{tabular}

\vspace{0.5cm}
\end{table}

\subsection{Model rejection potential optimisation for point source searches}
\label{pso}
\begin{figure}[htp]
\centering
 \mbox{\epsfig{file=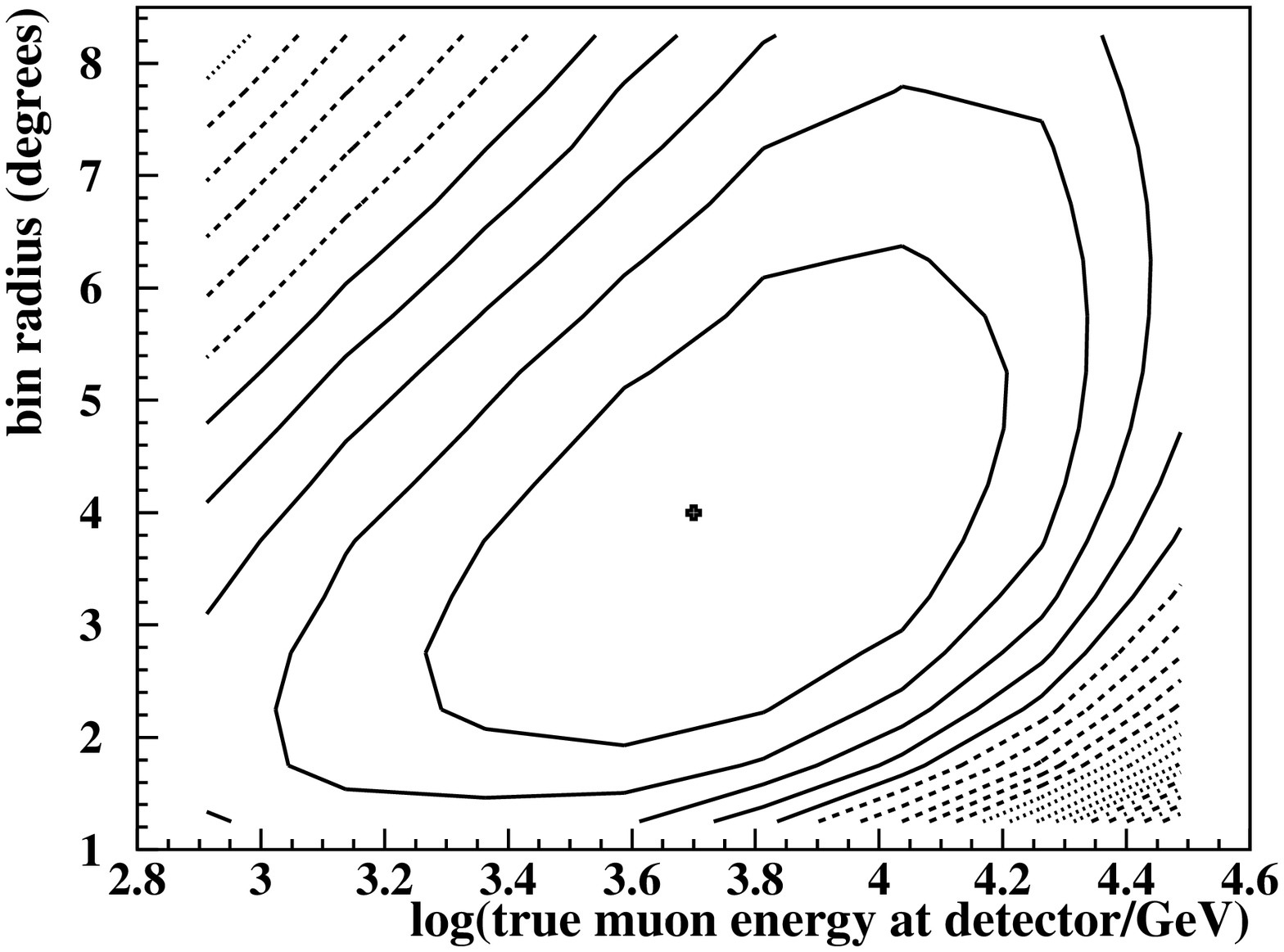,width=15cm}}
\caption[1]{
\label{ps_mrf_contour}Surface of model rejection factor, $\bar{\mu}/n_s$, for 
a point source search in a kilometre effective area neutrino telescope, where
the angular acceptance about the assumed source direction, and the true detector
muon energy are used as cut parameters. Adjacent contours correspond to a change in
the flux upper limit of $10^{-10}$,
 with the minimum value of $9.4 \times 10^{-10}\; \mathrm{GeV}\; \mathrm{cm}^{-2}\;
\mathrm{s}^{-1}\;$ indicated by the cross. 
 }
\end{figure}

In the search for a point source, we can reduce the background to the search by 
cutting on both the energy of the events, and also on the angular acceptance 
window about the known direction of the source.
 We apply the model
rejection potential procedure in two dimensions, producing a two dimensional
surface of model rejection factor, $\bar{\mu}/s$, in the energy and angular
acceptance ($\Delta\Psi$ degrees about the known direction)
 cut variables. As a single illustration of this technique, 
a contour plot representation
of this surface for the case of an $E^{-2}$ point source, assuming an angular
resolution with Gaussian $\sigma = 3.0^{\circ}$, and an exact measure of the
muon energy in the detector, is shown in figure~\ref{ps_mrf_contour}.
 The best constraint on the $E^{-2}$
point source (shown as a cross) is found by accepting
 events with $\Delta\Psi < 3.8^{\circ}$
 and $\log10(E_{\mu}/\mathrm{GeV}) > 4.5$. The remaining background of 1.4 events
leads to an average event upper limit of  3.56 and an
  optimised flux limit of 
\begin{equation}
 E^{2} \Phi_{90} < 9.4 \times 10^{-10}\; \mathrm{GeV}\; \mathrm{cm}^{-2}\;
\mathrm{s}^{-1}\; 
\end{equation}
The change in model rejection factor moving between adjacent contours corresponds
to a change in the average flux upper limit  in linear steps
of \mbox{$10^{-10}\; \mathrm{GeV}\; \mathrm{cm}^{-2}\;
\mathrm{s}^{-1}\;$}. Thus the minimum lies in a fairly broad range in the two 
dimensional cut parameter space. 

If we assume a 30\% logarithmic muon energy resolution, the optimal cuts change to 
 $\Delta\Psi < 2.5^{\circ}$
 and $\log10(E_{\mu}/\mathrm{GeV}) > 3.25$, and the limit is only a factor of 2.5 weaker
than for the true energy measurement case. Contrast this to the diffuse $E^{-2}$ limit case, where
the 30\% energy resolution produced a limit a factor of 4.9 weaker, due to sole reliance of
energy as the cut. In the point source case, much of the background rejection comes from the
angular acceptance cut, and the energy measurement error has less effect.

\subsection{Model rejection potential optimisation for gamma-ray burst searches}
The mechanisms that power gamma-ray bursts are still
not determined. If hadronic acceleration mechanisms are producing  gamma-rays 
via proton interactions and subsequent neutral pion decay, then there should also 
be fluxes of high energy neutrinos. Large neutrino detectors will look for
these neutrinos in coincidence with the satellite observations of the gamma-rays.
 Waxman and Bahcall~\cite{WBUB} compute the expected neutrino flux
by assuming that the  gamma-ray
bursts are also the source of the observed cosmic rays.

Just as in the point source search we optimise both the angular acceptance about the known 
source direction and the true muon energy. The additional cut on a 
 time window about the known
GRB time greatly reduces the background to the search. We assume that we will test
a model 
in which the neutrinos are produced within 10 seconds of the burst time. The
 accumulated number of coincident bursts is used as a
measure of the detector exposure~\cite{tyal}. 

\begin{figure}[htp]
\centering
 \mbox{\epsfig{file=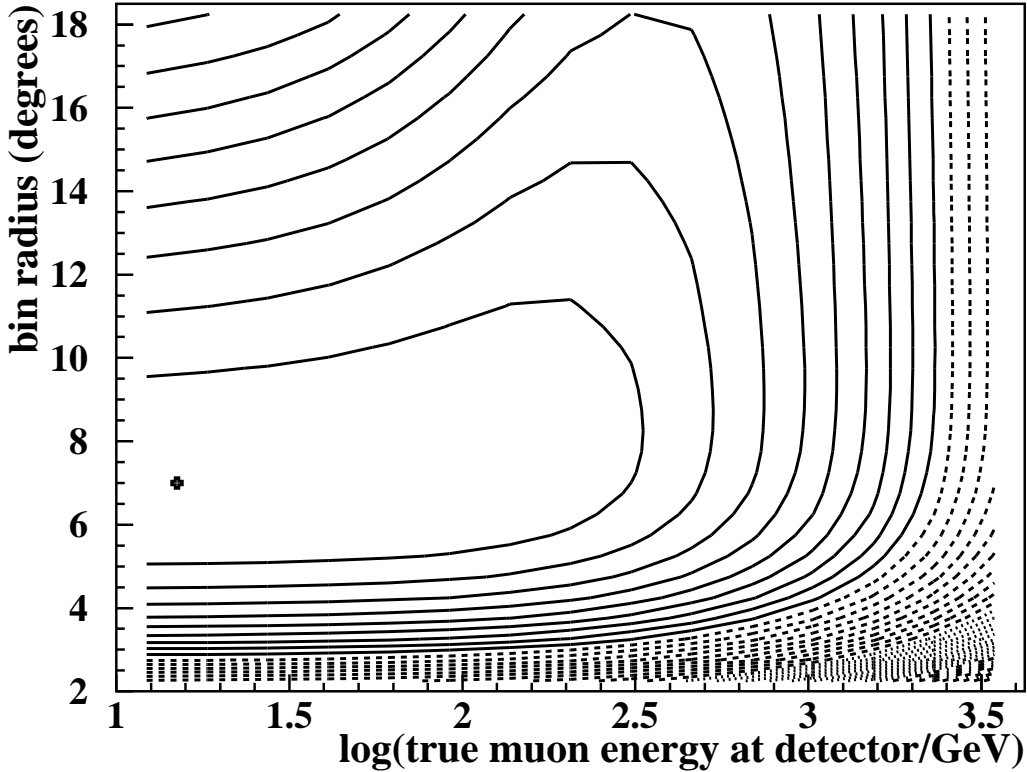,width=15cm}}
\caption[1]{
\label{grb_mrf_contour2}Surface of model rejection factor, $\bar{\mu}/n_s$, for 
a Waxman-Bahcall model
 gamma-ray burst source search in a kilometre effective area neutrino telescope, where, in
addition to angle and energy cuts, a time window cut of 10 seconds about each 
of the assumed 500 known GRB
emission times is applied. The point of the minimum model rejection factor of 0.18 is
indicated by the cross.}
\end{figure}

The model rejection factor surface after 500 satellite-coincident 
bursts are seen in the detector's field of view is shown in figure
\ref{grb_mrf_contour2}. The application of a 10 second time window cut 
about each of the known burst times reduces the total background to essentially zero
 and thus only loose cuts
are needed --
 an angular acceptance cut of  $\Delta\Psi < 7.0^{\circ}$
 and an energy cut of  $\log10(E_{\mu}/\mathrm{GeV}) > 1.18$, both of which are much weaker than
for the point source optimisation of section \ref{pso}, are found to be optimal.
 Using these cuts,
 the remaining atmospheric background (0.17 events)
and average upper limit (2.58) lead to an  optimal upper limit
corresponding to a flux of 0.18 times that of the Waxman-Bahcall prediction. Again, the minimum lies in
a broad region -- the  change in model
rejection factor  
between adjacent contours being $6.7 \times 10^{-3}$.
 Since the background is so close to zero, the limit will
tend to scale linearly with an increase in the number of observed bursts. 

Incorporating a realistic 30\% logarithmic  muon energy resolution 
leads to a model rejection factor of 0.3, again only a factor of 1.7 worse than the ideal muon
energy measurement case. The cuts stay close to those of the ideal case, with  $\Delta\Psi$ of 
$7.5^{\circ}$,
 and the energy cut loosened to accept all simulated events ($\log10(E_{\mu}/\mathrm{GeV}) > 1.0$) 
 leading to an expected background of 0.12 and average event
upper limit of 2.54. Due to the addition of the time cut, the energy measurement error has even
less weakening effect on the optimal limit than in the point source case.

\section{Conclusions} 
In this paper we have described  a method of choosing experimental cuts
in order to maximise the chance of placing
 the strongest possible constraint on an expected signal model. Choosing
the experimental cuts to optimise the model rejection factor (ratio of
expected average upper limit to expected signal) 
is shown to yield the best
average constraint on the signal model. We have demonstrated 
this method by
determining  the sensitivity of kilometre-scale neutrino detectors to
 diffuse and point sources of astrophysical neutrinos, and find that such 
detectors will strongly constrain present models. We suggest that 
the optimised  average flux upper limits described here be used as a standard
method of comparing the capabilities of different neutrino detectors.

\ack{We would like to thank Albrecht Karle, Ty DeYoung,
 Francis Halzen and David Steele for
useful discussions and comments on the manuscript, and Tom Gaisser 
for encouragement to bring this work to publication. We thank an anonymous
referee for general comments and for reminding us of the importance
 of the prompt charm contribution
to the atmospheric neutrino background. We acknowledge the 
support of the National Science Foundation, under contract number OPP-9980474.}

\clearpage

\end{document}